\documentclass[12pt,letterpaper]{article}
\usepackage[T1]{fontenc}
\usepackage[utf8]{inputenc}
\usepackage[margin=1in]{geometry}
\usepackage{amsmath,amsthm,amsfonts,amssymb,bm}
\usepackage{graphicx}
\usepackage[authoryear,round]{natbib}
\bibpunct{(}{)}{;}{a}{}{,}
\usepackage{booktabs,colortbl,float,makecell,dcolumn,multirow}
\usepackage{algorithm,algorithmicx,algpseudocode}
\usepackage{enumitem}
\usepackage{setspace}
\usepackage{microtype}
\usepackage{caption}
\usepackage[hidelinks]{hyperref}
\hypersetup{
  pdftitle={Statistical and Privacy Guarantees for Deep M-Regression via Stochastic Gradient Langevin Dynamics},
  pdfauthor={Yeheng Ge, Jian Huang, Shuyuan Wu},
  pdfkeywords={Deep neural network, Differential privacy, PAC-Bayesian theory, Stochastic gradient Langevin dynamics}
}
\theoremstyle{plain}
\newtheorem{theorem}{Theorem}

\newtheorem{corollary}{Corollary}
\newtheorem{lemma}{Lemma}
\newtheorem{assumption}{Assumption}
\theoremstyle{definition}

\newtheorem{remark}{Remark}
\newtheorem{definition}{Definition}
\floatname{algorithm}{Algorithm}

\newcommand{\bR}{\mathbb{R}}
\newcommand{\bE}{\mathbb{E}}
\newcommand{\mE}{\mathcal{E}}
\newcommand{\mD}{\mathcal{D}}
\newcommand{\mX}{\mathcal{X}}
\newcommand{\mY}{\mathcal{Y}}
\newcommand{\mB}{\mathcal{B}}
\newcommand{\mF}{\mathcal{F}}
\newcommand{\mH}{\mathcal{H}}
\newcommand{\mO}{\mathcal{O}}
\newcommand{\mW}{\mathcal{W}}
\newcommand{\KL}{\mathrm{KL}}
\newcommand{\mP}{\mathcal{P}}
\newcommand{\mQ}{\mathcal{Q}}

\title{\Large\bfseries Statistical and Privacy Guarantees for Deep M-Regression via Stochastic Gradient Langevin Dynamics}
\author{%
  Yeheng Ge$^{1}$ \quad Jian Huang$^{1,2}$\thanks{Corresponding author: Jian Huang, Department of Data Science and Artificial Intelligence, The Hong Kong Polytechnic University, Hung Hom, Kowloon, Hong Kong, China. Email: \href{mailto:j.huang@polyu.edu.hk}{j.huang@polyu.edu.hk}.} \quad Shuyuan Wu$^{3}$\\[0.4em]
  \parbox{0.95\textwidth}{\centering\small
    $^{1}$Department of Data Science and Artificial Intelligence,\\
    The Hong Kong Polytechnic University\\[0.2em]
    $^{2}$Department of Applied Mathematics,\\
    The Hong Kong Polytechnic University\\[0.2em]
    $^{3}$School of Statistics and Data Science,\\
    Shanghai University of Finance and Economics
  }
}
\date{}

\begin{document}
\begingroup
\singlespacing
\maketitle
\begin{abstract}
In this article, we study deep nonparametric regression trained via stochastic gradient Langevin dynamics (SGLD), a variant of stochastic gradient descent that injects carefully calibrated Gaussian noise during optimization.
Using a PAC--Bayesian analysis tailored to deep ReLU networks,
we establish an excess-risk bound that simultaneously accounts for network approximation error,  statistical complexity, and the algorithmic cost accumulated along the SGLD trajectory.
Under a Bernstein condition and within a stable SGLD regime, the excess-risk bound contains a stochastic term whose coefficient is inversely proportional to the variance of injected noise, while the estimator retains the optimal rate for nonparametric regression under the Hölder smoothness condition up to logarithmic factors. This result provides an approximation-error-aware, rate-optimal statistical perspective on the widely observed generalization benefits of moderate SGLD noise.
For per-sample clipped SGLD,
we further establish an exact
\(f\)-DP guarantee  and the corresponding asymptotic
\(\mu\)-GDP guarantee for the proposed estimator.
We establish statistical convergence rates and differential privacy guarantees for DNN estimators trained using SGLD, while  explicitly accounting for both network approximation error and the effect of injected noise in the excess-risk bound.
Extensive simulation studies and analyses of
two real datasets examine the finite-sample predictive performance and
privacy--utility behavior of the proposed procedure.

\end{abstract}
\par
\endgroup
\clearpage

\section{Introduction}

Classical statistical analysis typically treats noise as a nuisance: additional randomness increases variability and can degrade estimation accuracy. Deep learning presents a different perspective. Deep neural networks are trained over high-dimensional and highly non-convex loss surfaces, where the optimization path can directly affect the fitted model. Carefully injected noise can reshape this path, encourage exploration of low-risk regions, and lead to favorable stability and generalization behavior
\citep{mou2018generalization,li2019generalization,zhu2023uniform}.
Stochastic Gradient Langevin Dynamics (SGLD) is a canonical example. It augments each mini-batch stochastic-gradient update with isotropic Gaussian noise. The method retains the scalability of stochastic gradient descent and induces a distribution over network parameters,  which is  closely connected to approximate Bayesian posterior sampling
\citep{welling2011bayesian,teh2016consistency,li2016preconditioned}.
These features make SGLD attractive for deep regression and raise a fundamental statistical question: how does the injected Gaussian noise affect the generalization performance of the resulting estimator?

Answering this question is theoretically challenging. Existing statistical analyses of deep learning largely rely on uniform convergence tools, such as VC-dimension and Rademacher complexity bounds, which characterize the complexity of the entire function class without reference to the optimization path \citep{schmidt2020nonparametric,anthony2009neural}.
Such algorithm-agnostic bounds do not describe how injected noise shapes the trajectory of SGLD.
PAC--Bayesian analysis offers a natural alternative.
By adopting a Bayesian prior--posterior formulation, it provides generalization guarantees for randomized estimators in terms of their data-dependent parameter distributions, rather than the complexity of the entire function class.
\citet{mou2018generalization} and \citet{li2019generalization} applied PAC--Bayesian theory to noisy gradient methods in generic non-convex settings and derived data-dependent bounds through the Kullback--Leibler divergence between the algorithmic posterior and a reference prior; see also \citet{zhu2023uniform} and \citet{dadi2025generalization} for related developments. \citet{alquier2025minimax} developed PAC--Bayesian treatments of neural-network estimators for dependent data. What remains missing is a unified analysis that combines approximation error
and a fast-rate statistical bound
with an explicit characterization of the noise-level contribution for deep nonparametric regression.

Beyond its statistical role, Gaussian noise offers an important privacy benefit. Combined with per-sample gradient clipping and randomized mini-batch sampling, it can provide rigorous differential privacy  \citep{abadi2016deep,du2021dynamic,dong2022gaussian}.
For convex losses, \citet{chaudhuri2011differentially} formulated differentially private empirical risk minimization, \citet{bassily2014private} established tight error bounds and efficient algorithms for private convex optimization, and \citet{cai2021cost} derived minimax-optimal rates for private estimation in high-dimensional linear models.
Privacy--utility trade-offs have further been studied in kernel learning, inference on regression coefficients, multiple testing, and conformal inference
\citep{jain2013differentially,xia2025statistical,xia2023adaptive,zhang2025online}.
These studies provide a foundation for private
statistical learning. For deep neural networks, however, existing algorithm-agnostic learning theory does not directly link the statistical risk of SGLD-trained estimators to the privacy guarantees induced by clipping, subsampling, and Gaussian perturbation.
A unified theory is needed wherein privacy-preserving deep learning can be formally understood from the perspectives of statistical risk and privacy protection.

To address the above issues, we study deep nonparametric
\(M\)-estimation trained by SGLD and establish an excess-risk bound involving the approximation error, stochastic error, and
the Kullback--Leibler divergence between the algorithmic posterior of network weights and a data-independent reference prior. Our contributions are
summarized as follows.
\begin{itemize}
    \item First, we derive a  PAC--Bayes oracle
inequality for the
SGLD  posterior. The result is based on the Bernstein condition and leads to the fast statistical rate up to logarithmic factors. It explicitly involves the variance of injected Gaussian noise in the stochastic error term, helping to explain the generalization benefits of moderate SGLD noise.
    \item
Second, by combining this oracle inequality with ReLU approximation
results for H\"older classes, we show that, under the stated conditions,
the  estimator based on the SGLD posterior attains the optimal
nonparametric regression rate up to logarithmic factors.
\item
Third,
for
per-sample clipped SGLD with
randomized mini-batches
and heterogeneous step sizes,
we obtain an exact finite-sample
\(f\)-DP guarantee and the corresponding asymptotic
\(\mu\)-GDP guarantee.
    \item
Last, we
conduct simulation studies and real-data analyses to examine the
finite-sample prediction and privacy--utility performance of the
proposed procedure.
\end{itemize}

The remainder of the article is organized as follows. Section \ref{sec:method} introduces deep \(M\)-estimation, SGLD, and gradient clipping. Section \ref{sec:theory} establishes the excess-risk and privacy guarantees. Section \ref{sec:sim} presents the simulation studies, and Section \ref{sec:realdata} presents real-data applications. Section \ref{sec:con} concludes with a discussion of the main implications and limitations.

\section{Deep Nonparametric Regression with SGLD}
\label{sec:method}
This section develops the methodological framework. We first formulate the deep M-estimation problem and introduce the deep ReLU neural network architecture.
We then present SGLD.
Finally, we introduce per-sample gradient clipping, which bounds the sensitivity required for the privacy analysis in Section~\ref{sec:privacy}.

\subsection{Deep M-estimation Problems}

Suppose a random pair $Z=(X,Y)$ is drawn from an unknown joint
distribution $\mD$, with $X\in\mX\subset[0,1]^d$ and
$Y\in\mY\subset\bR$. For a loss function
$\ell:\mY\times\bR\to\bR$, write
$\ell(f,Z):=\ell\bigl(Y,f(X)\bigr)$.
In statistical learning, the target function $f^*$ is often defined as a population-risk minimizer:
\begin{equation}
\label{eq:population_risk}
f^*
\in
\operatorname*{arg\,min}_{f\in\mH}R(f),
\qquad
R(f)
=
\bE_{\mD}\big[\ell(f,Z)\big],
\end{equation}
where $\mH$ denotes the candidate function class.
The choice of the loss function $\ell(\cdot,\cdot)$ in \eqref{eq:population_risk} depends on the specific learning task at hand.
The framework includes the following familiar examples when $\mH$ contains the corresponding population target:
\begin{itemize}
\item Under the least-squares loss
$\ell(f,Z)=\frac{1}{2}\{Y-f(X)\}^2$,
the minimizer $f^*$ corresponds to the conditional expectation
$f^*(x)=\bE[Y\mid X=x]$, recovering classical nonparametric regression.
\item For binary classification with $Y\in\{-1,+1\}$, the logistic loss
$\ell(f,Z)=\log(1+\exp\{-Yf(X)\})$
yields the log-odds target
\[
f^*(x)
=
\log\left\{
\frac{\mathbb P(Y=1\mid X=x)}
     {\mathbb P(Y=-1\mid X=x)}
\right\}.
\]
\end{itemize}

Since the target function $f^*$ is unknown in practice, we employ a flexible function class to approximate it within the M-estimation framework. Deep neural networks, with their established approximation capabilities,
provide a natural nonparametric device for this purpose. Throughout, we focus on neural networks with the ReLU activation function $\sigma(t)=\max\{t,0\}$ \citep{yarotsky2017error,schmidt2020nonparametric}.
To be specific, consider a feedforward network $f_w$ with $L$ hidden layers and parameter vector $w$, and let \(p\) denote the number of
trainable parameters. We obtain
$f_w(x) = W_{L+1} \circ \sigma \circ W_{L} \circ \sigma \circ \cdots \circ W_{1}(x),$
where $W_l(\mathbf z)=\mathbf A_l\mathbf z+\mathbf b_l$ is the affine map at layer $l$, for $l=1,\ldots,L+1$, and $\sigma$ is the ReLU activation function applied componentwise.
The vector $w$ collects all the layer parameters $\{(\mathbf A_l,\mathbf b_l)\}_{l=1}^{L+1}$.

Subsequently, let
$\mF_n=\{f_w:w\in\mW\}$
denote the resulting ReLU network class. We define the population network oracle as
$$
f_{w^*}
=
\operatorname*{arg\,min}_{f\in\mF_n}R(f),
$$
which is indexed by the oracle parameter $w^*$.

\subsection{Stochastic Gradient Langevin Dynamics}

In practice, we observe an i.i.d.\ sample
$\mD_n= \{Z_1,\ldots,Z_n\}$, where $Z_i=(X_i,Y_i)$ are drawn from the population $\mD$.
Initialize $w_0\sim \mathcal N(0,\sigma_0^2\mathbf I_p)$, where $\sigma_0>0$ and $\mathbf I_p$ is the identity matrix of dimension $p$.
At iteration $k$, the SGLD update for the parameter vector $w_k$ is
\begin{equation}
\label{eq:noisy_sgd}
w_{k+1}
=
w_k
-
\eta_k\nabla_wR_{\mathcal B_k}(f_{w_k})
+
\sqrt{2\eta_k\gamma^{-1}}\,\xi_{k+1},
\qquad
\xi_{k+1}\sim\mathcal N(0,\mathbf I_{p}).
\end{equation}
 Here, $\{\eta_k\}_{k\ge0}$ is a positive decaying step-size
sequence.
 A canonical choice considered below is the polynomially
decaying schedule
$\eta_k=c_\eta(k+t_0)^{-\nu}$, where
$c_\eta, t_0>0$ and
$\nu\in(1/2,1)$.
The inverse-temperature parameter $\gamma>0$ is proportional to the inverse of the noise variance.
Smaller $\gamma$ corresponds to stronger injected noise
\citep{welling2011bayesian}. $\mathcal B_k$ is the set of sample indices in the $k$th mini-batch, sampled uniformly
from all subsets of $[n]$ of size $b$, and
$
R_{\mathcal B_k}(f_w)
=
b^{-1}
\sum_{i\in\mathcal B_k}
\ell(f_w,Z_i)
$
denotes the mini-batch
empirical risk. When $b=n$, we have $\mathcal B_k=[n]$ and
$
R_{\mathcal B_k}(f_w)
=
R_n(f_w)
=
n^{-1}
\sum_{i=1}^n
\ell(f_w,Z_i),
$
so \eqref{eq:noisy_sgd} reduces to the full-gradient
Langevin update. When $b=1$, it reduces to the
single-observation SGLD update.
 At non-differentiable points, $\nabla_w$ denotes a fixed measurable choice of subgradient.

To interpret the Gaussian
perturbation in \eqref{eq:noisy_sgd}, consider the
continuous-time Langevin diffusion associated with this
mean drift:
\begin{equation}
\label{eq:langevin_sde}
dw_t
=
-\nabla_wR_n(f_{w_t})\,dt
+
\sqrt{2\gamma^{-1}}\,dB_t,
\qquad
w_0\sim\mathcal N(0,\sigma_0^2\mathbf I_p),
\end{equation}
where $B_t$ is a standard $p$-dimensional Brownian motion.
Applying the Euler--Maruyama discretization to
\eqref{eq:langevin_sde} and replacing the full gradient
$\nabla_wR_n(f_{w_k})$ by its mini-batch estimate
$\nabla_wR_{\mathcal B_k}(f_{w_k})$ yields
\eqref{eq:noisy_sgd}. \citet{welling2011bayesian} used this
connection to introduce SGLD as a scalable method for
approximate Bayesian posterior sampling.
Subsequent
generalization analyses include
\citet{mou2018generalization}, \citet{li2019generalization},
\citet{zhu2023uniform}, and \citet{dadi2025generalization}.

In the context of deep learning, the loss landscape has many local minima of vastly differing generalization properties.
The moderate Gaussian perturbations help the iterates
to persist in a broad low-risk neighborhood, rather than remaining in a narrow local minimum.
Consequently, noisy training may favor locally stable solutions with better generalization ability.

\subsection{Gradient Clipping}\label{subsec:clipping}

In addition to the potential optimization benefit described above,
Gaussian perturbation can be calibrated to provide differential
privacy once the sensitivity of each update is bounded. We therefore
replace the raw mini-batch gradient
$\nabla_wR_{\mathcal B_k}(f_{w_k})$ in
\eqref{eq:noisy_sgd} by an average of per-sample clipped gradients.
For a clipping threshold $\tau>0$, define
\begin{equation}\label{eq:clipped_grad}
\begin{aligned}
\bar g_i(w)
=
\frac{\nabla_w\ell(f_w,Z_i)}
{\max\!\left\{
1,
\|\nabla_w\ell(f_w,Z_i)\|_2/\tau
\right\}},\quad
g_k(w,\mathcal B_k)
=
\frac{1}{b}
\sum_{i\in\mathcal B_k}
\bar g_i(w).
\end{aligned}
\end{equation}
By construction,
$
\|\bar g_i(w)\|_2\le\tau$
and $\|g_k(w,\mathcal B_k)\|_2\le\tau.
$
The resulting clipped SGLD update is
\begin{equation}\label{eq:noisy_sgd_clipped}
w_{k+1}
=
w_k
-
\eta_k g_k(w_k,\mathcal B_k)
+
\sqrt{2\eta_k\gamma^{-1}}\,\xi_{k+1},
\qquad
\xi_{k+1}\sim\mathcal N(0,\mathbf I_{p}).
\end{equation}
Algorithm~\ref{alg:sgld_privacy} summarizes the resulting clipped SGLD procedure with randomized mini-batch sampling and Gaussian perturbation.
The resulting privacy guarantees are established in Section~\ref{sec:privacy}.

 Gradient clipping changes the update drift and hence the distribution
of the optimization trajectory, but the resulting output law remains
amenable to PAC--Bayesian analysis under the stated loss assumptions.
PAC-Bayesian analysis allows general gradient
estimators and controls their contribution through their norms \citep{london2017pac,mou2018generalization}.
In our
setting, the change-of-measure argument therefore uses the actual clipped
updates; it does not require them to be unbiased estimates of the raw
gradient.

This formulation provides flexibility beyond analyses tailored to linear
models with specific tail conditions. For example,
\citet{cai2021cost} study private linear regression with Gaussian errors
and bounded-design assumptions, using Gaussian tail control to calibrate
response truncation.
A similar sub-Gaussian condition is considered by \citet{xia2025statistical}.
The path-KL bound used below instead requires no
sub-Gaussian tail assumption on the raw stochastic gradients and applies
equally to nonlinear, non-convex network models.

\begin{algorithm}[!htbp]
\caption{Clipped SGLD}\label{alg:sgld_privacy}
\begin{algorithmic}[1]
\Require data $\mathcal{D}_n=\{Z_i\}_{i=1}^n$, loss $\ell$,
initial variance $\sigma_0^2>0$, step sizes
$\{\eta_k\}_{k=0}^{K-1}$,  inverse-noise parameter
$\gamma$, clipping threshold $\tau$, and batch size $b$.
\State Draw
$w_0\sim\mathcal N(0,\sigma_0^2\mathbf I_{p})$.
\For{$k=0,\ldots,K-1$}
\State Draw
$\mathcal B_k\sim
\mathrm{Unif}\{B\subset[n]:|B|=b\}$
independently of previous draws.
\For{$i\in\mathcal B_k$}
    \State Compute $\bar g_i(w_k)$ as in
    \eqref{eq:clipped_grad}.
\EndFor
\State Set
$g_k(w_k,\mathcal B_k)
=
b^{-1}\sum_{i\in\mathcal B_k}\bar g_i(w_k)$.
\State Draw
$\xi_{k+1}\sim\mathcal N(0,\mathbf I_{p})$
independently.
\State Update $w_{k+1}$ as in
\eqref{eq:noisy_sgd_clipped}.
\EndFor
\Ensure The learned network weights
$(w_1,\ldots,w_K)$.
\end{algorithmic}
\end{algorithm}

\section{Theoretical Results}
\label{sec:theory}

We now develop statistical theory for the SGLD estimator within the deep \(M\)-estimation framework.
We first introduce the notation used below.
For two non-negative sequences $\{a_n\}$ and $\{b_n\}$, we write $a_n = \mO(b_n)$ if there exists a constant $C>0$ such that $a_n \le C b_n$ for all sufficiently large $n$, and $a_n = \widetilde{\mO}(b_n)$ if $a_n = \mO(b_n \log^c n)$ for some constant $c \ge 0$.
For a sequence of random variables $\{X_n\}$, we write $X_n=\mO_{\mathbb P}(b_n)$ if $X_n/b_n$ is bounded in probability, and $X_n=\widetilde{\mO}_{\mathbb P}(b_n)$ if $X_n=\mO_{\mathbb P}(b_n\log^c n)$ for some constant $c\ge0$.
We write $a_n = o(b_n)$ if $a_n/b_n \to 0$.
We use $\bE$ for expectation, $\bR$ for the real line,
$\KL$ for Kullback--Leibler divergence,
and the parameter space is denoted by $\mW$.

\subsection{Excess Risk Bound}

\noindent
Recall that \(\mD_n=(Z_1,\ldots,Z_n)\) represents the training sample, where
\(Z_i=(X_i,Y_i)\) are independent copies of \(Z=(X,Y)\sim\mD\).
For a
network parameter \(w\in\mW\subset\bR^{p}\), let \(f_w\) be the
ReLU network and define
$
R(f_w)=\bE_{\mD}\ell(f_w,Z)
$ and
$R_n(f_w)=\frac1n\sum_{i=1}^n\ell(f_w,Z_i)$ as the population and empirical risks, respectively.
The population and empirical excess risks are
\[
\mE(f_w)=R(f_w)-R(f^*),
\qquad
\mE_n(f_w)=R_n(f_w)-R_n(f^*).
\]
For a probability measure \(\rho\) on the parameter space, we use the shorthand
\[
\begin{aligned}
\rho[\mE]&=\int \mE(f_w)\,d\rho(w),\quad
\rho[\mE_n]&=\int \mE_n(f_w)\,d\rho(w).
\end{aligned}
\]
For $k=0,\ldots,K$, let $\mQ_k^\gamma=\mathcal L(w_k\mid \mD_n)$ denote the conditional law of the SGLD iterate \(w_k\). The algorithm-averaged excess risk is
\[
\mQ_K^\gamma[\mE(f_w)]
=
\int \mE(f_w)\,d\mQ_K^\gamma(w)
=
\int\{R(f_w)-R(f^*)\}\,d\mQ_K^\gamma(w).
\]
Equivalently, $\mQ_K^\gamma[\mE(f_w)]=\bE\{R(f_{w_K})-R(f^*)\mid \mD_n\}$, where the expectation is over all randomness used by SGLD.   Before establishing the formal
results, we first state the regularity conditions used in the analysis.

\begin{assumption}[Bernstein condition]
\label{ass:bernstein}
For every \(w\in\mW\), define
$
h_w(Z)=\ell(f_w,Z)-\ell(f^*,Z)-\mE(f_w).
$
There exist constants \(K_B,C_B>0\), independent of $n$, $p$, and $w$,
 such that, for every integer \(m\ge2\),
\[
\bE_{\mD}|h_w(Z)|^m
\le
\frac{m!}{2}V_w C_B^{m-2},
\qquad
V_w\le K_B\,\mE(f_w).
\]
\end{assumption}
A Gaussian reference law on $\bR^p$ is compatible with
Assumption~\ref{ass:bernstein}, which controls loss fluctuations at each
fixed parameter. Uniformly bounded predictors can be obtained through
output truncation or a bounded output map, such as $T_M\circ g_w \text{ or }
M\tanh(g_w),$ where the truncation map is defined by
$T_M(u)=\max\{-M,\min(M,u)\}$ for $M>0$.
These constructions constrain the predictions while allowing
the parameters to range over $\bR^p$. When applied as post-processing,
they leave the SGLD recursion and the Gaussian reference law unchanged. In Section~A.3 of the Supplementary Material, we verify the Bernstein condition for squared, quantile, and Huber losses and provide sufficient conditions under which it holds.

Then we give the definition of the H\"older class.
\begin{definition}[$\beta$-Hölder class]
Let $\beta>0$ and write
$
    \beta=k+a,   k\in\mathbb{N}_0,    a\in(0,1],
$
where, when $\beta$ is an integer, we adopt the convention
$k=\beta-1$ and $a=1$.
For $B>0$, the $\beta$-Hölder class
$\mathcal{H}^{\beta}([0,1]^d,B)$ consists of all functions
$f\in C^k([0,1]^d)$ satisfying
$
    \max_{\lvert\alpha\rvert\leq k}
    \left\|D^\alpha f\right\|_{\infty}
    \leq B
$
and
$
    \max_{\lvert\alpha\rvert=k}
    \sup_{\substack{x,y\in[0,1]^d\\x\neq y}}
    \frac{
        \left|D^\alpha f(x)-D^\alpha f(y)\right|
    }{
        \|x-y\|_2^a
    }
    \leq B.
$
Here, $\alpha=(\alpha_1,\ldots,\alpha_d)\in\mathbb{N}_0^d$ is a
multi-index,
$
    \lvert\alpha\rvert
    =\sum_{j=1}^d\alpha_j,
$$
    D^\alpha f
    =
    \frac{
        \partial^{\lvert\alpha\rvert}f
    }{
        \partial x_1^{\alpha_1}\cdots
        \partial x_d^{\alpha_d}
    },
$
and
$
    \|D^\alpha f\|_\infty
    =
    \sup_{x\in[0,1]^d}|D^\alpha f(x)|.
$
\end{definition}

\begin{assumption}[Network approximation]\label{ass:approx}
Suppose the target function \(f^*\) belongs to the H\"older space
\(\mH^\beta([0,1]^d,B)\). Let
\(\mF_n(L_n,W_n,p,B_n)\) denote the class of ReLU networks with depth
at most \(L_n\), width at most \(W_n\), at most \(p\) trainable coordinates, and parameter magnitudes bounded by \(B_n\). There exists a
sequence \(\epsilon_n\to0\) such that, for all sufficiently large
\(n\), the class \(\mF_n\) contains a network \(f\) satisfying
\[
\|f-f^*\|_\infty\le C_1\epsilon_n
\qquad\text{and}\qquad
\mE(f)\le C_2\epsilon_n^2.
\]
The network structure satisfies
\(L_n=\mO\{\log(1/\epsilon_n)\}\),
\(W_n=\mO(\epsilon_n^{-d/\beta})\),
\(p=\mO\{\epsilon_n^{-d/\beta}\log(1/\epsilon_n)\}\), and
\(B_n=\mO(\epsilon_n^{-\xi})\), where \(C_1\) and \(\xi\) are
positive constants depending only on \(d,\beta\), and \(B\).
\end{assumption}

Assumption~\ref{ass:bernstein} imposes a Bernstein moment bound on the
centered loss and links its variance to the excess risk.
Closely related Bernstein conditions are standard in fast-rate
PAC--Bayesian analysis; see
\citet{mhammedi2019pac} and \citet{grunwald2021pac}.

Assumption~\ref{ass:approx} requires the ReLU class to approximate the
H\"older target at rate \(\epsilon_n\) in the uniform norm.
For squared loss, the excess-risk bound follows from the
uniform approximation bound, whereas other losses require an
appropriate curvature, margin, or calibration condition.
Similar ReLU
approximation conditions are standard in deep nonparametric regression;
see \citet{yarotsky2017error} and \citet{schmidt2020nonparametric}.
The parameter bound $B_n$ is imposed only on a deterministic
approximating network. It does not restrict the Gaussian reference law or
the SGLD iterates, which are defined on $\bR^p$. When $\|f_{w^*}\|_\infty< M$ is bounded, output truncation preserves the uniform approximation error and can
be implemented by a fixed ReLU module of constant size.

{
To relate the empirical and population excess risks of the
data-dependent SGLD output, we use a PAC--Bayesian inequality
\citep{catoni2007home}. Under Assumption~\ref{ass:bernstein},
for any data-independent reference measure $\mP$, with probability
at least $1-\delta$, simultaneously for all $\mQ$ with finite
$\KL(\mQ\|\mP)$,
\[
\mQ[\mE]
\lesssim \mQ[\mE_n]
+\frac{\KL(\mQ\|\mP)+\log(1/\delta)}{n}.
\]
This uniform guarantee is also applicable to the
data-dependent output law $\mQ_K^\gamma$.

To connect the KL term to the SGLD trajectory, we choose the
terminal law of a data-independent reference process with the
same initialization and Gaussian noise as SGLD but no gradient drift:
\[
v_0\sim\mathcal N(0,\sigma_0^2I_p),\qquad
v_{k+1}=v_k+\sqrt{2\eta_k/\gamma}\,\zeta_{k+1},
\]
where the $\zeta_{k+1}$ are independent standard Gaussian vectors,
independent of $v_0$ and the training sample.
For the prescribed step sizes and noise parameter, this gives
\[
\mP_{\mathrm{ref},K}:=\mathcal L(v_K)
=\mathcal N\!\left(
0,\left\{\sigma_0^2+\frac{2T_K}{\gamma}\right\}I_p
\right),
\qquad T_K=\sum_{k=0}^{K-1}\eta_k.
\]
The reference therefore provides a baseline for measuring how the accumulated gradient updates change the output distribution.
This construction isolates the contribution of the gradient drift
to the KL complexity, following the trajectory-based analyses of
\citet{mou2018generalization,li2019generalization}.

Combining the PAC--Bayesian bound with the network approximation
condition and the empirical optimization gap yields the following
oracle inequality.
}

\begin{lemma}[Oracle inequality for the SGLD  posterior]
\label{lem:posterior-excess}
Suppose Assumptions~\ref{ass:bernstein} and \ref{ass:approx}  hold.  Then, for every
\(\delta\in(0,1)\), with probability at least \(1-2\delta\),
\begin{align}
\label{eq:main-excess-bound}
\mQ_K^\gamma[\mE( f_w)]
\lesssim
\mE( f_{w^*})
+\varepsilon_{\mathrm{opt},K}
+\frac{1}{n}\{\KL(\mQ_K^\gamma\|\mP_{\mathrm{ref},K})+\log(1/\delta)\},
\end{align}
where
$\varepsilon_{\mathrm{opt},K} = \int R_n(f_w)\,d\mQ_K^\gamma(w)
-
\inf_{u\in\mW}R_n( f_u)$
is the nonnegative empirical optimization gap for the
predictors.
\end{lemma}
Lemma~\ref{lem:posterior-excess} gives a fast-rate oracle inequality,
with a statistical complexity term of order
$\{\KL(\mQ_K^\gamma\|\mP_{\mathrm{ref},K})\}/n$.
This fast-rate scaling relies on the Bernstein condition in
Assumption~\ref{ass:bernstein}.
The KL divergence in \eqref{eq:main-excess-bound} arises from the
PAC--Bayes change of measure from the data-independent reference law
$\mP_{\mathrm{ref},K}$ to the data-dependent SGLD output law
$\mQ_K^\gamma$. It serves as a statistical complexity
term when converting empirical excess risk into population excess
risk.

We next bound the KL term in Lemma~\ref{lem:posterior-excess}.
Conditional on $\mD_n$, let $\mathbb Q_{0:K}$ and $\mathbb P_{0:K}$
be the SGLD and reference path laws.
For the reference process constructed above, the KL chain rule
expresses the path KL as a sum of expected KL divergences between
the parameter transitions. Given the same current iterate and
mini-batch, the SGLD and reference transitions, denoted by
$Q_k(\cdot\mid w_k,\mathcal B_k)$ and $P_k(\cdot\mid w_k)$,
are Gaussian
with the same covariance and means differing by the gradient update.
The terminal parameter contains no more information for distinguishing
the two processes than the full path. Thus, by the data-processing
inequality, its KL is no greater than the path KL. Combining these
steps gives
\[
\begin{aligned}
\KL(\mQ_K^\gamma\|\mP_{\mathrm{ref},K})
&\le\KL(\mathbb Q_{0:K}\|\mathbb P_{0:K})=\sum_{k=0}^{K-1}\bE\!\left[
\KL\!\left\{Q_k(\cdot\mid w_k,\mathcal B_k)
\|P_k(\cdot\mid w_k)\right\}\mid\mD_n\right]\\
&=\frac{\gamma}{4}\sum_{k=0}^{K-1}\eta_k\,
\bE\!\left[\|g_k(w_k,\mathcal B_k)\|_2^2\mid\mD_n\right]
\le\frac{\gamma p}{4}G_K,
\end{aligned}
\]
where $\|u\|_2^2\le p\|u\|_\infty^2$ and
\[
G_K=\sum_{k=0}^{K-1}\eta_k\,
\bE\!\left[\|g_k(w_k,\mathcal B_k)\|_\infty^2\mid\mD_n\right].
\]
Thus the KL penalty is bounded by a term of order $\gamma pG_K/n$.
We next control the network approximation error $\mE(f_{w^*})$ and
the statistical complexity.

In the following analysis,
we
focus exclusively on the statistical error and the
network approximation error $\mE(f_{w^*})$.
Controlling the optimization error $\varepsilon_{\mathrm{opt},K}$ is beyond the scope of this paper
and is not pursued here.

\begin{remark}
In this paper, we do not analyze the optimization error and assume that it is sufficiently small not to affect the optimal convergence rate.
    Specifically,
    if
$\varepsilon_{\mathrm{opt},K}
\lesssim p/n$,
the target approximation and stochastic errors are the leading terms for excess risk.
In practical implementations under the training
schedules considered here, SGLD is run until the empirical objective
has stabilized, and the terminal iterates typically remain close in
empirical risk to the minimizer
\citep{raginsky2017nonconvex,teh2016consistency}.
The sampled-estimator results reported in  Section C.2 of the Supplementary Materials provide additional
empirical support for stability across the terminal output distribution. Intuitively, the injected noise can help SGLD escape narrow minima
and explore wider valleys of low empirical risk. This interpretation
is related to the wide-valley geometry studied by
\citet{chaudhari2017entropy} and \citet{keskar2017large}.
\end{remark}

\begin{theorem}[Conditional optimal rate for H\"older regression]
\label{thm:minimax}
Suppose Assumptions~\ref{ass:bernstein} and \ref{ass:approx} hold and the optimization error is of the same order as, or smaller than, the statistical and approximation errors and therefore does not change the convergence rate. Let $f^*\in\mH^\beta([0,1]^d,B)$. Choose
$\epsilon_n\asymp n^{-\beta/(2\beta+d)}$,
$p=\widetilde{\mO}\{n^{d/(2\beta+d)}\}$,
and a ReLU architecture satisfying
Assumption~\ref{ass:approx}.
 Then
\[
\mQ_K^\gamma[\mE(f_w)]
=
\widetilde\mO
\bigl(
n^{-2\beta/(2\beta+d)}
\bigr)
+
\widetilde\mO_{\mathbb P}\bigl(
\gamma G_K n^{-2\beta/(2\beta+d)}
\bigr).
\]
If $\gamma G_K=O_{\mathbb P}((\log n)^c)$ for some $c>0$, we have
\[
\mQ_K^\gamma[\mE(f_w)]
=
\widetilde\mO_{\mathbb P}\bigl(
n^{-2\beta/(2\beta+d)}
\bigr).
\]
 \end{theorem}
The first term,
\(\widetilde{\mO}\{n^{-2\beta/(2\beta+d)}\}\), is the cost of approximating  a  \(\beta\)-H\"older regression function
with ReLU neural networks
\citep{yarotsky2017error,schmidt2020nonparametric}.
The second term $\widetilde\mO_{\mathbb P}\bigl(\gamma G_K n^{-2\beta/(2\beta+d)}\bigr)$ is the algorithm-dependent stochastic error.
Since the injected-noise variance is proportional
to \(\gamma^{-1}\), increasing the Gaussian perturbation corresponds
to possibly reducing this
stochastic error term.
Theorem~\ref{thm:minimax} requires $\gamma G_K=O_{\mathbb P}((\log n)^c)$ for some $c>0$ to obtain the stated rate.
Theorem~\ref{thm:minimax} clarifies how injected noise enters the excess-risk bound:
stronger noise reduces the explicit factor $\gamma$ in the stochastic term,
while also shaping $G_K$ and $\varepsilon_{\mathrm{opt},K}$ through the SGLD trajectory.
When these trajectory-dependent quantities remain controlled, this result is
consistent with the observed generalization benefits of moderate noise in SGLD,
while preserving the optimal statistical rate under the stated conditions.

Furthermore, we
consider the realized estimator \(f_{w_K}\), where
\(w_K\) is the terminal iterate with conditional law \(w_K\mid \mD_n\sim\mQ_K^\gamma\), and give the following high-probability
result.

\begin{corollary}[Statistical guarantee for the realized terminal predictor]
\label{cor:sampled-estimator}
Under the assumptions of Lemma~\ref{lem:posterior-excess}, let \(w_K\)
denote the terminal iterate generated by SGLD, so that
\(w_K\mid \mD_n\sim\mQ_K^\gamma\). Then, for any \(\alpha \in(0,1)\), with probability at least $1-2\delta-\alpha$ over the sample and algorithmic randomness,
\[
\mE(f_{w_K})
\lesssim
\frac{1}{\alpha}
\left\{
\mE(f_{w^*})
+
\frac{\gamma pG_K}{n}
+\varepsilon_{\mathrm{opt},K} + \frac{\log(1/\delta)}{n}
\right\}.
\]
Hence, under the conditions of
Theorem~\ref{thm:minimax}, the realized terminal network also satisfies
\[
\mE(f_{w_K})
=
\widetilde{\mO}_{\mathbb P}\!\left(
 n^{-2\beta/(2\beta+d)}
\right).
\]
In particular, for any fixed \(\alpha\in(0,1)\), the realized terminal
iterate has the same statistical order as the algorithm-averaged
bound.
\end{corollary}

\begin{remark}[From the output distribution to the realized iterate]
Conditional on the training sample,
$
\bE\{\mE(f_{w_K})\mid \mD_n\}
=
\mQ_K^\gamma[\mE].
$
Since the excess risk is non-negative, Markov's inequality implies
$
\mathbb P
\left\{
\mE(f_{w_K})
>
\alpha^{-1}\mQ_K^\gamma[\mE]
\mid  \mD_n
\right\}
\le\alpha.
$
Corollary~\ref{cor:sampled-estimator} therefore converts the
algorithm-averaged PAC--Bayes guarantee into a tail bound for the
terminal network actually produced by SGLD. A smaller \(\alpha\)
provides a stronger probability guarantee at the cost of the factor
\(1/\alpha\). For any fixed \(\alpha\), this factor does not alter the
asymptotic order.
\end{remark}
\begin{corollary}[Statistical guarantee for clipped SGLD]
\label{cor:clipped-sgld}
Let \(w_K^{(C)}\) denote the terminal iterate generated by the clipped
SGLD in Algorithm~\ref{alg:sgld_privacy},  and define
$
\mQ_K^{(C),\gamma}
=
\mathcal L(w_K^{(C)}\mid\mD_n).
$
Let $g_k^{(C)}$ denote the clipped mini-batch gradient defined in \eqref{eq:clipped_grad}, and let
$
G_K^{(C)}
=
\sum_{k=0}^{K-1}\eta_k
\bE\big[\|g_k^{(C)}(w_k^{(C)},\mathcal B_k)\|_\infty^2
\mid\mD_n\big]
$.
Define the empirical optimization gap by
\[
\varepsilon_{\mathrm{opt},K}^{(C)}
=
\int R_n(f_w)\,d\mQ_K^{(C),\gamma}(w)
-
\inf_{u\in\mW}R_n(f_u).
\]
We assume that this optimization error satisfies the corresponding condition in Theorem~\ref{thm:minimax}.
Suppose Assumption~\ref{ass:bernstein} and the corresponding conditions of Theorem~\ref{thm:minimax} hold, with \(G_K\)
replaced by \(G_K^{(C)}\). Then
\[
\mQ_K^{(C),\gamma}[\mE(f_w)]
=
\widetilde{\mO}\!\left(
n^{-2\beta/(2\beta+d)}
\right)
+
\widetilde{\mO}_{\mathbb P}\!\left(
 n^{-2\beta/(2\beta+d)}
\right).
\]
Thus, clipped SGLD has the same statistical order as the unclipped
algorithm-averaged bound whenever its cumulative-gradient quantity and
empirical optimization gap satisfy the corresponding conditions.
\end{corollary}

\begin{remark}[Effect of adding an $\ell_2$ regularizer]
\label{rem:l2-regularization-decay}
For the regularized objective
$
R_n(f_w)+\frac{\lambda_{\mathrm{reg}}}{2}\|w\|_2^2,
$  the SGLD recursion becomes
$
w_{k+1}
=
(1-\eta_k\lambda_{\mathrm{reg}})w_k
-\eta_k g_k(w_k,\mathcal B_k)
+\sqrt{2\eta_k\gamma^{-1}}\,\xi_{k+1}.
$
Adding the penalty $\lambda_{\mathrm{reg}}\|w\|_2^2/2$ introduces
the shrinkage factor $1-\eta_k\lambda_{\mathrm{reg}}$ into the SGLD update.
This motivates an exponentially weighted analogue of $G_K$,
schematically written as
\[
\sum_{k=0}^{K-1}\eta_k
\exp\{-c\lambda_{\mathrm{reg}}(T_K-T_k)\}
\bE\!\left[\|g_k(w_k,\mathcal B_k)\|_\infty^2\mid\mD_n\right],
\]
where $c>0$ and $T_k=\sum_{j=0}^{k-1}\eta_j$;
see \citet{mou2018generalization} and \citet{li2019generalization}.
By discounting earlier gradient contributions, this weighting offers
a mechanism for controlling the trajectory-dependent complexity.
For bounded $\gamma$, a corresponding trajectory term of at most
polylogarithmic order preserves the optimal statistical rate up to
logarithmic factors, provided the optimization error remains controlled.
\end{remark}

\subsection{Differential Privacy Guarantee}\label{sec:privacy}

We next establish the privacy guarantee for the proposed SGLD algorithm with gradient clipping in Algorithm~\ref{alg:sgld_privacy}.
The statistical and privacy results apply to the same clipped
SGLD algorithm and therefore provide simultaneous guarantees
under their combined assumptions; see
Corollary~\ref{cor:joint-guarantees}.
The privacy guarantee itself does not require the
cumulative-gradient or optimization-gap conditions.
The statistical rate remains conditional on these additional
conditions, which are not shown to follow automatically
from the privacy calibration.

\begin{definition}[\(f\)-DP and GDP, \citet{dong2022gaussian}]
\label{def:fdp-gdp}
(i) Under replacement adjacency, two datasets $D$ and $D'$, each of size $n$, are adjacent if one record in $D$ is replaced to obtain $D'$.
(ii) For two probability laws \(\mathsf P_1\) and \(\mathsf P_2\) on the same measurable space, their trade-off function is
\[
T(\mathsf P_1,\mathsf P_2)(\alpha)
=\inf_{0\le \phi\le 1}
\left\{
1-\int \phi\,d\mathsf P_2:
\int \phi\,d\mathsf P_1\le \alpha
\right\},
\qquad 0\le \alpha\le 1,
\]
where the infimum is taken over all measurable tests.
(iii) A mechanism \(\mathcal M\) is \(f\)-DP if \(T\{\mathcal L(\mathcal M(D)),\mathcal L(\mathcal M(D'))\}\ge f\) for every adjacent pair \((D,D')\).
(iv) Let \(\Phi\) denote the standard normal distribution function and define \(\mathcal G_\mu(\alpha)=\Phi\{\Phi^{-1}(1-\alpha)-\mu\}\) for \(\mu\ge0\). A mechanism is \(\mu\)-Gaussian differentially private, or \(\mu\)-GDP, if it is \(\mathcal G_\mu\)-DP.
\end{definition}

\begin{definition}[Asymptotic GDP]
\label{def:asymptotic-gdp}
A sequence of mechanisms \(\{\mathcal M_n\}_{n\ge1}\) is said to admit an asymptotic \(\mu\)-GDP guarantee if there exist symmetric trade-off functions \(\{F_n\}_{n\ge1}\) such that \(\mathcal M_n\) is \(F_n\)-DP for every \(n\) and \(\|F_n-\mathcal G_\mu\|_\infty\to0\), where the norm is the supremum norm on \([0,1]\).
\end{definition}

For a trade-off function \(f\), let \(\mathrm{Id}(\alpha)=1-\alpha\) and
$
f^{-1}(\alpha)=\inf\{t\in[0,1]: f(t)\le \alpha\}.
$
When a fixed-size mini-batch is sampled at rate \(q=b/n\) and the selected indices are not released, the \(f\)-DP subsampling operator is
$
C_q(f)=\{\min(f_q,f_q^{-1})\}^{**}$ for  $
f_q=qf+(1-q)\mathrm{Id},
$
where \(h^{**}\) denotes the greatest convex lower bound of \(h\).
For trade-off functions \(f=T(\mathsf P_1,\mathsf P_2)\)
and \(g=T(\mathsf Q_1,\mathsf Q_2)\), define their tensor product by
$
f\otimes g
:=
T(\mathsf P_1\otimes\mathsf Q_1,\,
  \mathsf P_2\otimes\mathsf Q_2),
$
where \(\mathsf P_i\otimes\mathsf Q_i\) denotes the product
probability measure for \(i=1,2\), and
\(\bigotimes_{k=0}^{K-1}f_k\) denotes its repeated application.
Throughout the asymptotic analysis below, the number of iterations \(K\),
the batch size \(b\), the clipping threshold \(\tau\),
the noise parameter \(\gamma\), and the step-size schedule
\(\{\eta_k\}_{k=0}^{K-1}\) may depend on \(n\).
We impose the following assumption.

\begin{assumption}
\label{ass:exact-heterogeneous-gdp}
Assume that
$
b/n\to0
\text{ and }
\sup_n\max_{0\le k<K} \tau \sqrt{2\gamma\eta_k}/b<\infty,
$
where $b$ is the batch size and $\tau$ is the clipping threshold.
\end{assumption}
Assumption~\ref{ass:exact-heterogeneous-gdp} requires the mini-batch sampling fraction \(b/n\) to vanish and the iteration-specific Gaussian privacy parameters \(\tau\sqrt{2\gamma\eta_k}/b\) to remain uniformly bounded; a closely related asymptotic regime is considered in \citet{dong2022gaussian}. The following theorem first establishes the
exact finite-sample \(f\)-DP composition and then, under this
assumption, derives its asymptotic GDP limit.

\begin{theorem}[Privacy guarantee]
\label{thm:fdp_privacy}
(i)
Algorithm~\ref{alg:sgld_privacy} satisfies $f$-DP with
\[
f =
\bigotimes_{k=0}^{K-1}
C_{b/n}\!\left(\mathcal G_{\tau\sqrt{2\gamma\eta_k}/b}\right).
\]
(ii)
Furthermore, under Assumption~\ref{ass:exact-heterogeneous-gdp}, suppose that
$
n^{-2}b^2\sum_{k=0}^{K-1}
\chi\!\left(\tau\sqrt{2\gamma\eta_k}/b\right)
\longrightarrow \mu^2/2
$
for some \(\mu>0\), where \(\chi(u)=e^{u^2}\Phi(3u/2)+3\Phi(-u/2)-2\) for \(u\ge0\). Then Algorithm~\ref{alg:sgld_privacy} admits an asymptotic \(\mu\)-GDP guarantee in the sense of Definition~\ref{def:asymptotic-gdp}.
\end{theorem}
We next specialize the
general condition to two commonly used step-size schedules: polynomial
decay and terminal decay, and provide directly verifiable sufficient
conditions in Corollary~\ref{cor:quadratic-gdp}.

\begin{corollary}
\label{cor:quadratic-gdp}
Suppose Assumption~\ref{ass:exact-heterogeneous-gdp} holds.
\begin{itemize}
\item[(i)] Suppose \(\eta_k=c_\eta(k+t_0)^{-\nu}\), where \(c_\eta>0\), \(t_0>0\), and \(\nu\in(0,1)\). If \(\tau^2\gamma c_\eta K^{1-\nu}/\{(1-\nu)n^2\}\to\mu^2/2\) for some \(\mu>0\), then Algorithm~\ref{alg:sgld_privacy} admits an asymptotic \(\mu\)-GDP guarantee.
\item[(ii)] Suppose \(\eta_k=\eta_0(1-k/K)^\nu\), where \(\eta_0>0\) and \(\nu>0\). If \(\tau^2\gamma\eta_0K/\{(\nu+1)n^2\}\to\mu^2/2\) for some \(\mu>0\) and \((\tau/b)\sqrt{2\gamma\eta_0}\to0\), then Algorithm~\ref{alg:sgld_privacy} admits an asymptotic \(\mu\)-GDP guarantee.
\end{itemize}
More generally, for arbitrary positive step sizes \(\{\eta_k\}_{k=0}^{K-1}\), the same conclusion holds if
\[
\frac{\tau^2\gamma}{n^2}\sum_{k=0}^{K-1}\eta_k\to\frac{\mu^2}{2}
\quad\text{for some }\mu>0,
\qquad
\frac{\tau^3\gamma^{3/2}}{bn^2}\sum_{k=0}^{K-1}\eta_k^{3/2}\to0.
\]
\end{corollary}

\begin{remark}
The preceding asymptotic GDP guarantees can also be interpreted on
the more familiar \((\varepsilon,\delta)\)-DP scale through the
standard conversion below.
By \citet[Corollary~1]{dong2022gaussian}, \(\mathcal G_\mu\)-DP is equivalent to \((\varepsilon,\delta)\)-DP for every \(\varepsilon\ge0\), with
\[
\delta=
\Phi\!\left(-\frac{\varepsilon}{\mu}+\frac{\mu}{2}\right)
-e^\varepsilon\Phi\!\left(-\frac{\varepsilon}{\mu}-\frac{\mu}{2}\right).
\]
\end{remark}

\begin{remark}[Comparison with conventional DP--SGD]
\label{rem:sgld-vs-noisy-sgd-privacy}
Compared   with equation \eqref{eq:noisy_sgd}, conventional
DP--SGD \citep{dong2022gaussian} can be written as
\[
w_{k+1}
=
w_k-\eta_k g_k(w_k,\mathcal B_k)
+
\eta_k\sqrt{2\gamma^{-1}}\,
\xi_{k+1},
\qquad
\xi_{k+1}\sim\mathcal N(0,\mathbf I_p).
\]
In conventional DP--SGD, the standard deviation of the Gaussian perturbation is proportional to $\eta_k$, whereas in SGLD it is proportional to $\sqrt{\eta_k}$. Thus the privacy accounting analysis above for SGLD is different from that in the conventional DP-SGD algorithm. To be specific, the composition result of \citet{dong2022gaussian} therefore gives
$
\mu_{\mathrm{DP\text{-}SGD}}^2
\sim
2\tau^2\gamma K/n^2
$
when $\tau\sqrt{\gamma}/b\to0$.
For clipped SGLD,  Corollary~\ref{cor:quadratic-gdp}
 gives
$
\mu_{\mathrm{SGLD}}^2
\sim
 2\tau^2\gamma
\sum_{k=0}^{K-1}\eta_k/n^2.
$
Thus, conventional DP--SGD accumulates privacy loss according to the
raw iteration count $K$, whereas SGLD is governed by the cumulative
step size $\sum_k\eta_k$.
For example, let
$\eta_k=c_\eta(k+t_0)^{-\nu}$ with $\nu\in(1/2,1)$, so that
$\sum_{k<K}\eta_k\asymp K^{1-\nu}$. If
$b\asymp\sqrt n$, $\tau\asymp\sqrt{\log n}$, then
$\tau\sqrt{\gamma}/b\to0$, and maintaining the same fixed nonzero
asymptotic GDP level gives
\[
K_{\mathrm{DP\text{-}SGD}}
\asymp
\frac{n^2}{\log n},
\qquad
K_{\mathrm{SGLD}}
\asymp
\left(
\frac{n^2}{\log n}
\right)^{1/(1-\nu)}.
\]
Therefore, under these calibrations, the privacy constraint permits
substantially more SGLD iterations than conventional DP--SGD at the
same asymptotic GDP level.
\end{remark}

Building on the preceding analysis, we obtain the following joint statistical and privacy guarantees.
\begin{corollary}[Joint statistical and privacy guarantees]
\label{cor:joint-guarantees}
Consider the clipped SGLD algorithm in
Algorithm~\ref{alg:sgld_privacy}.
Suppose Assumptions~\ref{ass:bernstein} and
\ref{ass:approx} hold, with the architecture chosen as in
Theorem~\ref{thm:minimax}. Assume further that
$
\varepsilon_{\mathrm{opt},K}^{(C)}
\lesssim p/n$ so that the optimization error is of the same order as the target statistical rate or smaller.
If
\[
\gamma G_K^{(C)}
=O_{\mathbb P}\{(\log n)^c\},
\]
for some fixed $c\ge0$, then
\[
\mathcal Q_K^{(C),\gamma}[\mathcal E]
=
\widetilde O_{\mathbb P}
\bigl(n^{-2\beta/(2\beta+d)}\bigr).
\]
The same algorithm simultaneously satisfies the finite-sample
$f$-DP guarantee in Theorem~\ref{thm:fdp_privacy}(i).
If, additionally, the conditions of
Theorem~\ref{thm:fdp_privacy}(ii) hold, it admits the
stated asymptotic $\mu$-GDP guarantee.
\end{corollary}

A comparison of the scope of our theoretical results with
related studies is provided in Table~S1 of the Supplementary Material.

\section{Simulation}
\label{sec:sim}

In this section, we examine the finite-sample performance of SGLD.
We first present a toy example to illustrate how moderate Gaussian
perturbation may be helpful for out-of-sample prediction.
We then conduct
more systematic experiments for deep \(M\)-regression and clipped
SGLD, focusing respectively on generalization across noise levels and the resulting privacy--utility trade-off.  Further implementation details are provided in
Section~C of the Supplementary Materials.
The test losses summarize predictive risk across the reported configurations. Throughout this section, $\gamma^{-1}$ denotes the injected-noise scale, with parameter-noise covariance $2\eta_k\gamma^{-1}I_p$ at update $k$. The baseline $\gamma^{-1}=0$ removes this Gaussian perturbation while retaining mini-batch sampling.

\subsection{Illustrative Example}
\label{subsec:toy-examples}

\textbf{Example 1: Sparse Linear Regression with LASSO and SCAD.}
We first consider a sparse linear model.
The linear class contains the regression function, so its approximation error is zero. This example illustrates how prediction and estimation errors vary with injected noise in the presence of penalty bias, finite-time optimization error, and sampling variability.

The data are generated from
\[
y_i=x_i^\top w^*+\varepsilon_i,\qquad
\varepsilon_i\sim\mathcal N(0,0.5^2),\qquad i=1,\ldots,n,
\]
with $(n,d)=(1200,60)$, where $d$ is the covariate dimension.
The covariates are Gaussian with an AR(1) covariance matrix $\Sigma$, where $\Sigma_{jl}=0.70^{|j-l|}$ for $j,l=1,\ldots,d$. The coefficient vector has two active pairs,
$
(w^*_{2k-1},w^*_{2k})=(5,-5), k=1,2,
$
and all remaining coordinates are zero.
The regularized mini-batch gradient $\nabla \widehat{R}_t(w)$ is computed using mini-batch $\mB_t$ and the corresponding objective function is
\[
\widehat{R}_t(w)
=
\frac{1}{|\mB_t|}\sum_{i\in\mB_t}\frac{1}{2}(y_i-x_i^\top w)^2
+\sum_{j=1}^d P(w_j),
\]
where $P(w_j)$ is either the convex LASSO penalty or the non-convex SCAD penalty.

\begin{figure}[!htbp]
\centering
\includegraphics[width=0.8\linewidth]{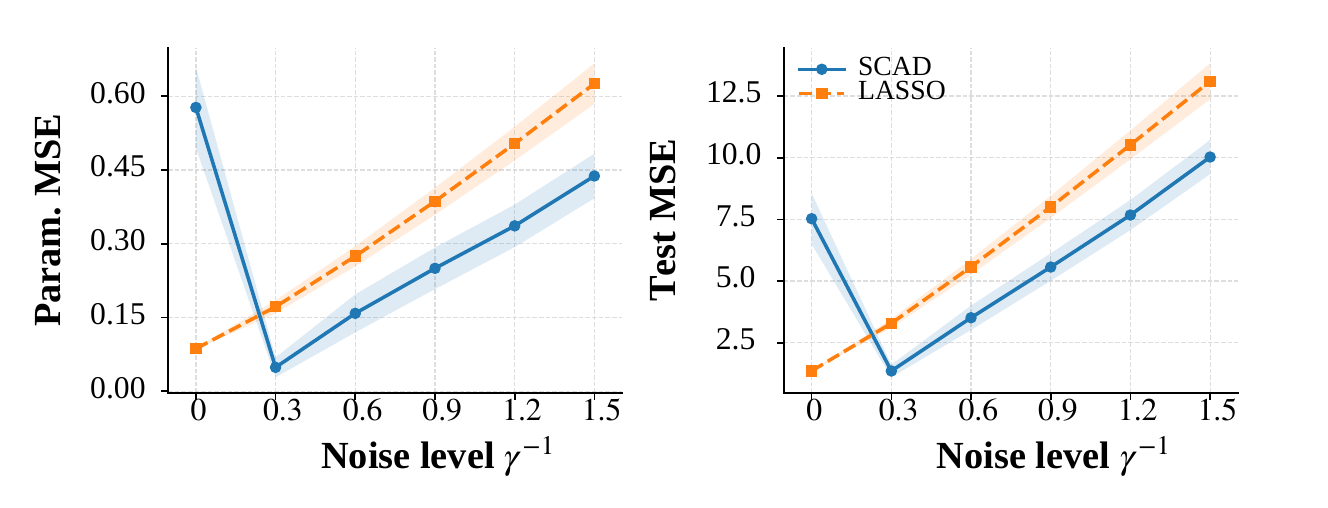}
\caption{Sparse linear regression with SCAD and LASSO penalties. The left panel reports the tail-averaged parameter MSE, and the right panel reports the tail-averaged test MSE, across the injected-noise levels $\gamma^{-1}$.}
\label{fig:linear-penalty-error}
\end{figure}

Figure~\ref{fig:linear-penalty-error} shows a clear difference between
the SCAD and LASSO settings. For the non-convex SCAD penalty, both the
parameter MSE and test MSE attain their smallest values at
\(\gamma^{-1}=0.3\), and both are lower than their noise-free
counterparts. At this tested noise level, the displayed estimation and prediction errors are lower than those of the baseline. For the convex LASSO
penalty, both errors are minimized at \(\gamma^{-1}=0\), and the positive noise levels tested yield larger errors.

\begin{figure}[!htbp]
\centering
    \includegraphics[width=\linewidth]{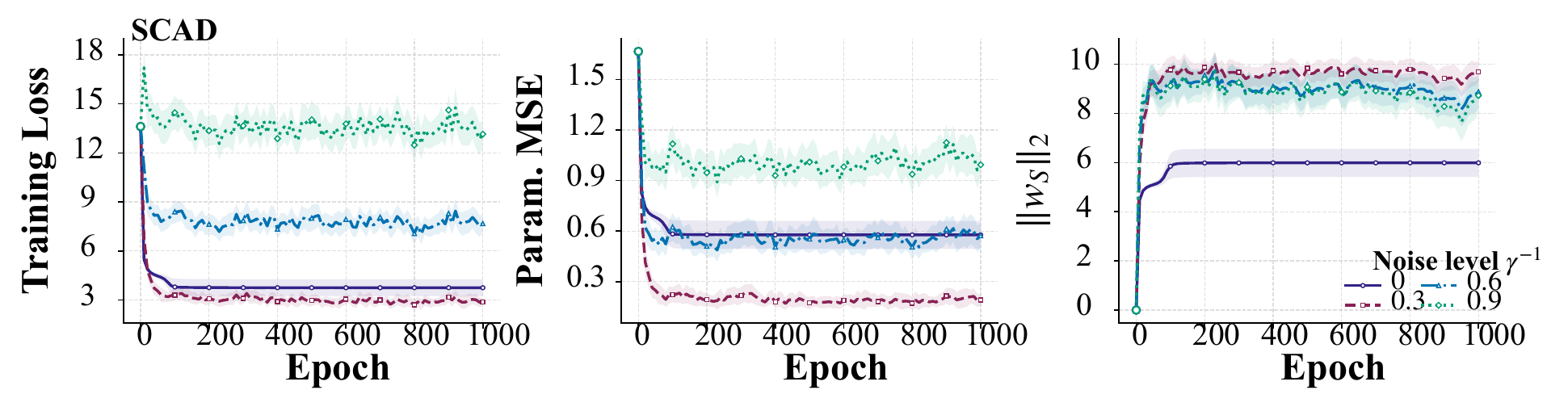}
    \includegraphics[width=\linewidth]{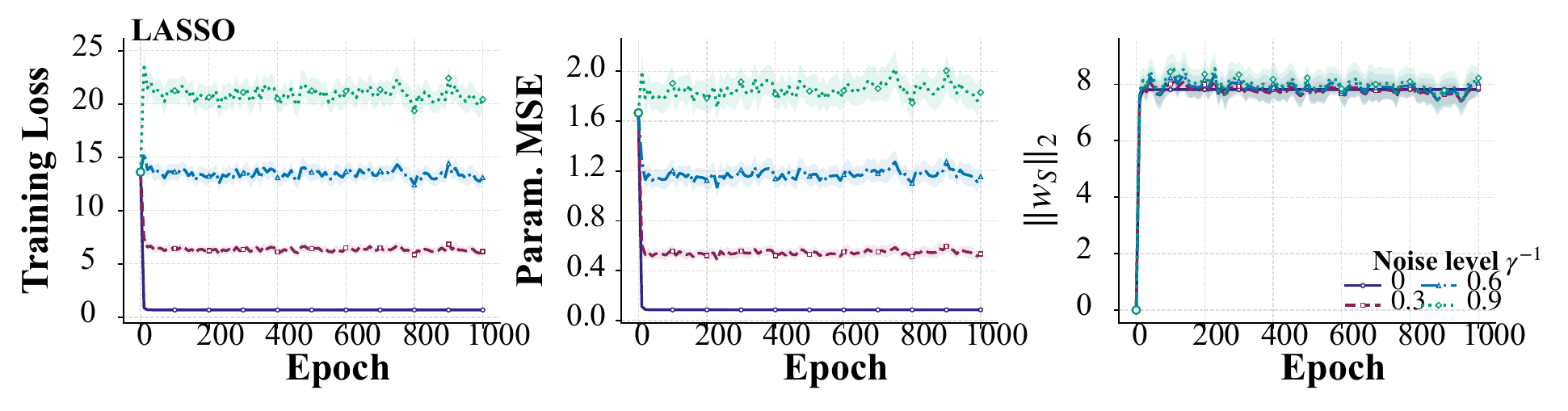}
\caption{Optimization trajectories under the SCAD (upper panel) and LASSO (lower panel) penalties. From left to right, each row displays the current loss value, current parameter MSE, and the $\ell_2$ norm of the active coordinates.
The true active-coordinate norm is $10$.
}
\label{fig:linear-penalty-path}
\end{figure}

Figure~\ref{fig:linear-penalty-path} illustrates the associated training dynamics. Under SCAD, the noise-free trajectory remains in a region with relatively large parameter error. The active-coordinate
norm remains near \(6\), well below the true value \(10\), and the
parameter MSE remains relatively large. With moderate Gaussian
perturbation, the trajectory leaves this region, moves toward the
correct active-coordinate scale, and reaches a solution with lower
parameter error. In contrast, the LASSO objective is convex and has no
suboptimal local minima. Its noise-free trajectory already reaches a
stable solution, while additional noise increases both the objective
value and the parameter error. The displayed SCAD trajectories are consistent with a potential optimization benefit of moderate perturbation.

\subsection{Generalization Performance in Multivariate Deep
Regression}

We next conduct a more systematic study of the generalization
performance of SGLD in multivariate deep regression. We vary the
sample size, input dimension, and injected-noise level, and evaluate
out-of-sample performance over 100 independent replications.
Section C of the Supplementary Materials contains the implementation details and simulation results for additional quantile levels and the Huber loss.

The experiments use a fully connected network with two hidden layers
of width 32. The network is trained for 800
epochs using mini-batches of size 50.
The
predictions are averaged over the checkpoints from the final \(10\%\)
of training. Writing $\mathcal T$ for the retained checkpoints, the predictor is $\bar f(x)=|\mathcal T|^{-1}\sum_{k\in\mathcal T}f_{w_k}(x)$, and test losses are evaluated using $\bar f$. The reported means and standard deviations summarize variation across replications; the comparisons below describe these observed patterns.

\textbf{Example 2: Quantile regression.}
Let \(\widetilde X\sim N(0,\Sigma)\), where
\(\Sigma_{jk}=0.2^{|j-k|}\), and define
\(X_j=4\{\Phi(\widetilde X_j)-1/2\}\). We consider the
heteroscedastic location--scale model
\[
Y=m(X)+s(X)\epsilon,
\qquad
\epsilon\sim N(0,1),
\]
where
\[
\begin{aligned}
m(x) ={}& 3x_1+1.5(2x_2+1)(2x_3-1)+0.2\sin(\pi x_4)\\
        &+0.4\cos(\pi x_4)+({2-\sin(\pi x_5)})^{-1}2\sin(\pi x_5),\\
s(x) ={}& 1+0.3x_5(|x_6|+1)^2.
\end{aligned}
\]
The conditional quantile is therefore
$
Q_q(Y\mid X=x)
=
m(x)+|s(x)|\Phi^{-1}(q).
$
We estimate the conditional quantile at \(q=0.2\) using the check loss.

\begin{figure}[!htbp]
\centering
\includegraphics[width=0.9\linewidth]
{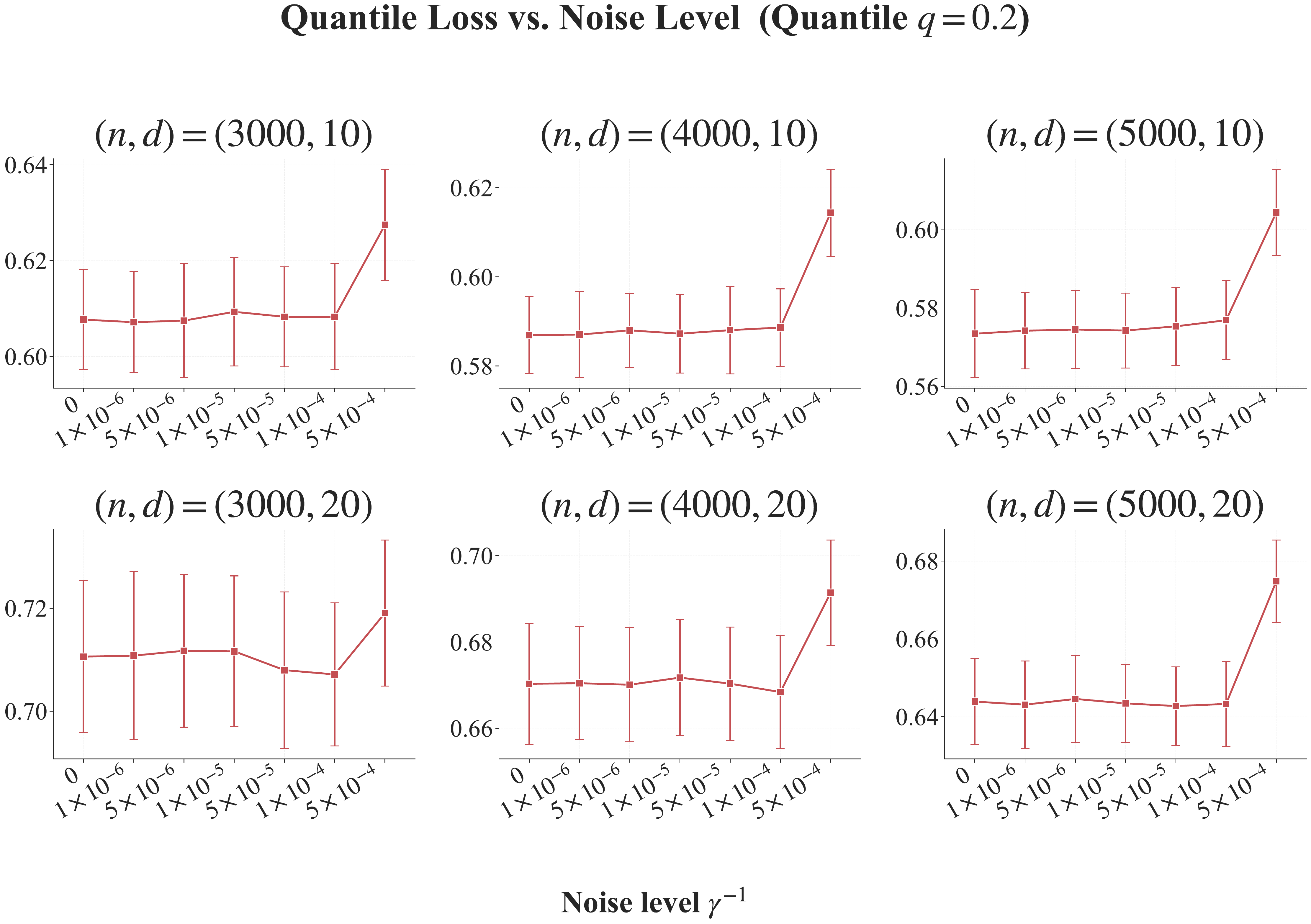}
\caption{Test quantile loss for quantile regression at \(q=0.2\)
under different noise levels. Points and error bars report the mean
and standard deviation, respectively, on an independent test set of
size 10,000 over 100 independent replications.}
\label{fig:generalization-quantile}
\end{figure}

Figure~\ref{fig:generalization-quantile} shows a quantitatively
consistent pattern across all six combinations of sample size \(n\)
and input dimension \(d\). At small and moderate noise levels, the mean
test quantile losses are comparable to, and in several settings
slightly lower than, their noise-free counterparts. When the injected-noise level increases to
\(\gamma^{-1}=5\times10^{-4}\), the test quantile loss exceeds the
noise-free value in all six settings. The mean losses vary relatively little across the smaller-noise settings, with some below the baseline; the largest tested noise level gives higher mean losses in every panel.
Results for \(q=0.5\) and \(q=0.8\) are presented in Section C of the
Supplementary Materials and show a similar pattern.
 Together, these results illustrate how the observed prediction errors vary across the tested noise levels.

We also present results for the sampled estimator at quantile level $q=0.2$ in Section C.2 of the Supplementary Materials. In this example, the sampled estimator has empirical test losses similar to those of the checkpoint-averaged predictor reported in the main text.

\subsection{Privacy--Utility Performance of Clipped SGLD}

We finally examine the privacy--utility performance of clipped SGLD
in a multivariate least-squares regression setting.

\textbf{Example 3: Least-squares regression.}
Let \(\widetilde X\sim N(0,\Sigma)\), where
\(\Sigma_{jl}=0.2^{|j-l|}\) for $j,l=1,\ldots,d$, and define
\(X_j=4\{\Phi(\widetilde X_j)-1/2\}\). The response is generated from
\[
Y=m(X)+\epsilon,
\qquad
\epsilon\sim N(0,1),
\]
where
$
m(x)
=
3x_1
+
1.5(2x_2+1)(2x_3-1)
+
2x_5(|x_6|+1)^2.
$
The covariate distribution and network architecture are the same as those
in Example 2. We set \(d=20\) and consider
\(n\in\{3000,4000,5000\}\). Training uses mini-batches of size
\(b=\lfloor\sqrt n\rfloor\) and
\(K=\lfloor0.02n^2/\log n\rfloor\) parameter updates. At each update,
the per-sample gradients are clipped as in
\eqref{eq:clipped_grad}, with threshold
\(\tau_n=2\sqrt{\log n}\). We consider
$
\gamma^{-1}
\in
\left\{
0,\,
10^{-5},\,
5\times10^{-5},\,
10^{-4},\,
5\times10^{-4}
\right\}.
$
SGLD models from the
final \(10\%\) of the training updates are averaged for prediction.
For each positive noise level, we report the
asymptotic GDP parameter in
Theorem~\ref{thm:fdp_privacy}(ii).

\begin{figure}[!htbp]
\centering
\includegraphics[width=0.9\linewidth]{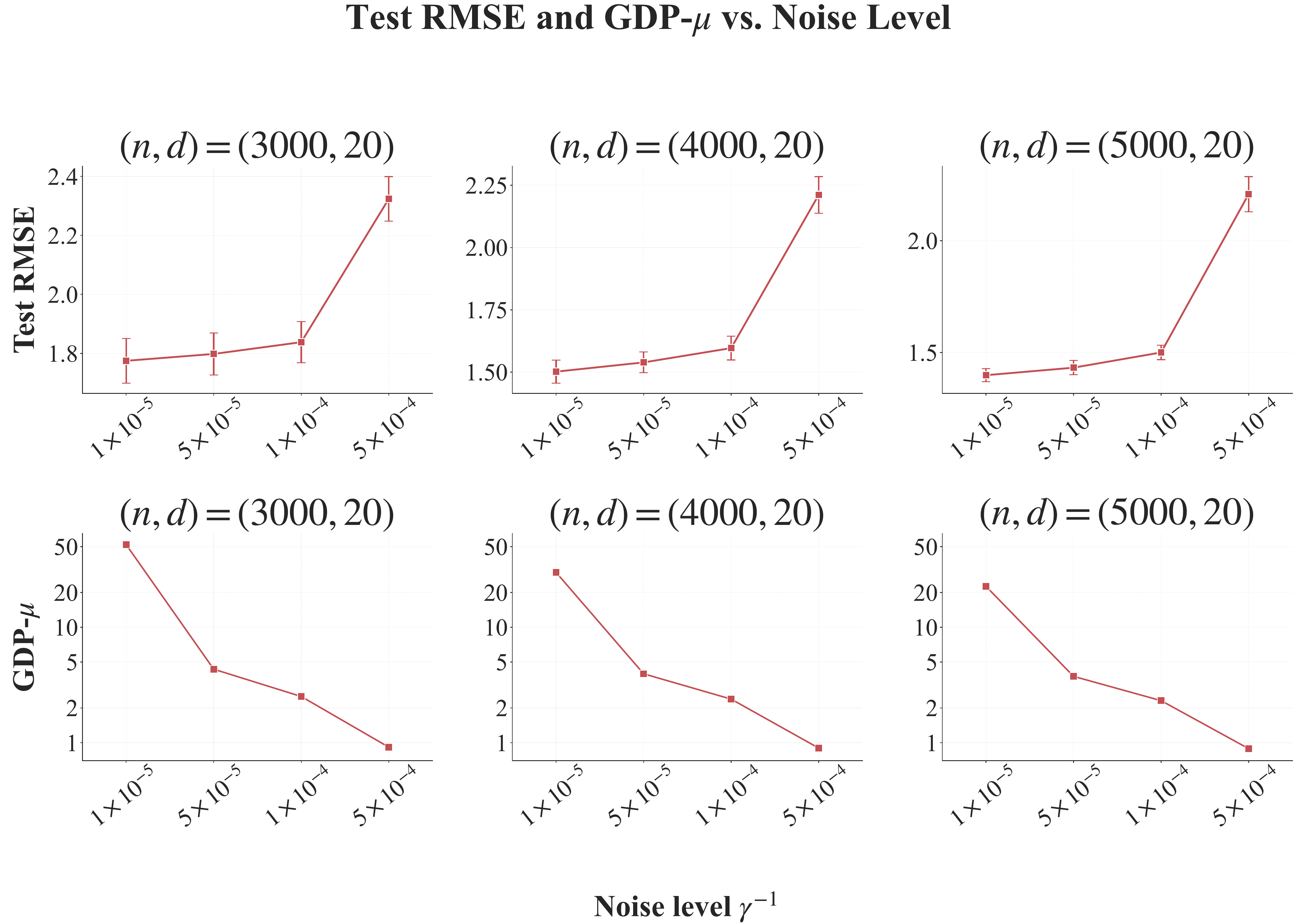}
\caption{Prediction and privacy performance of least-squares regression
with per-sample gradient clipping. The upper panels report the mean
test RMSE, with error bars showing the mean \(\pm\) one standard
deviation over 100 independent replications.
}
\label{fig:clipped-lse-gdp}
\end{figure}
Figure~\ref{fig:clipped-lse-gdp} shows a similar privacy--utility pattern
across the three sample sizes. At \(\gamma^{-1}=10^{-4}\), the mean test
RMSE is \(1.838\), \(1.596\), and \(1.500\) for \(n=3000\), \(4000\),
and \(5000\), respectively. Increasing the noise level to
\(\gamma^{-1}=5\times10^{-4}\) reduces \(\mu\), but raises
the mean test RMSE above \(2.2\) in all three settings. The mean test RMSE changes less over the smaller-noise settings and is higher at the largest tested noise level, alongside a decrease in the reported asymptotic GDP parameter.

\section{Real Data Analysis}
\label{sec:realdata}
In this section, we apply the proposed method to the publicly available
ACTG-175 and MNIST datasets, which represent regression and
classification tasks, respectively.
Both analyses use the clipped update in
Algorithm~\ref{alg:sgld_privacy}.
For both datasets, we report the asymptotic GDP parameter $\mu$
corresponding to Theorem~\ref{thm:fdp_privacy}(ii).  We compare the reported mean test metrics across noise levels to describe the accompanying changes in predictive performance and the asymptotic privacy parameter.

\subsection{The ACTG-175 Dataset}
\label{subsec:actg175}

We first consider the AIDS Clinical Trials Group Study 175
(ACTG-175) dataset from \citet{hammer1996trial}, available through the
R package \textit{speff2trial}.\footnote{\url{https://search.r-project.org/CRAN/refmans/speff2trial/html/ACTG175.html}}
The original study was a randomized, double-blind clinical trial that
compared four antiretroviral treatment regimens for adults infected
with human immunodeficiency virus type 1. The dataset contains records
from 2,139 patients and 27 variables. These variables describe
demographic characteristics, treatment history, baseline immunologic
measurements, randomized treatment assignments, and follow-up
outcomes. The patient-level clinical records provide a natural setting
for privacy-preserving prediction \citep{suriyakumar2021chasing}.

We use the  CD4 T-cell count measured at approximately 20 weeks as the
response, denoted by \(Y\). The predictors include five continuous
variables: age, body weight, Karnofsky score, baseline CD4 count, and
baseline CD8 count. We also include seven binary baseline indicators
for hemophilia, homosexual activity, intravenous drug use, race,
gender, treatment-history stratum, and symptomatic status, together
with three indicators for randomized treatment assignment. The
resulting input dimension is 15. In each replication, 90\% of the
observations are randomly assigned to the training set and the
remaining 10\% to the test set. The split is stratified by treatment
arm, giving 1,925 training observations and 214 test observations.

We fit a one-hidden-layer ReLU network with width 32 and 545 trainable
parameters. Training uses mini-batches of size 64, and each
per-sample gradient is clipped at \(\tau=4.859<2\sqrt{\log(1,925)}\).
We set
\(\eta_k=10^{-3}(1-k/K)^{0.501}\). Final predictions are obtained by
averaging the predictions from network states collected during the
last 20\% of the training trajectory. We compute
\[
\operatorname{NLL}_{\rm test}
=
\frac{1}{n_{\rm test}}
\sum_{i=1}^{n_{\rm test}}
\left\{
\frac{1}{2}\log(2\pi\widehat\sigma^2)
+
\frac{\{Y_i-\bar f(X_i)\}^2}{2\widehat\sigma^2}
\right\},
\]
where $  \widehat\sigma^2 =  \sum_{i\in\mathcal I_{\rm train}} \{Y_i-\bar f(X_i)\}^2 / n_{\rm train} $  is the mean squared residual of the tail-averaged predictor on the corresponding training set.
The NLL is used solely for internal evaluation and does not enter the training objective or the SGLD updates. Neither the NLL nor the residual variance is included in the private release, so this internal evaluation incurs no additional privacy cost.
We also report test RMSE.
The experiment is repeated
independently 20 times.

\begin{figure}[!htbp]
\includegraphics[width=0.6\linewidth]
{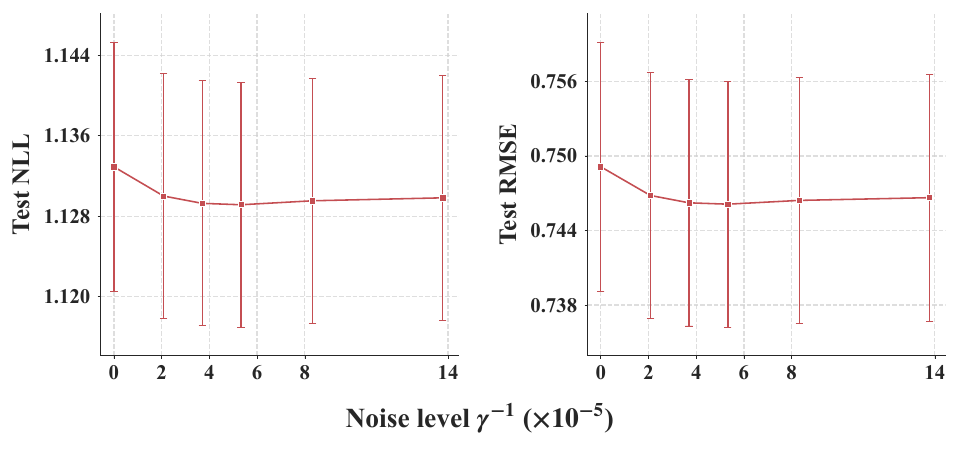}
\includegraphics[width=0.33\linewidth]
{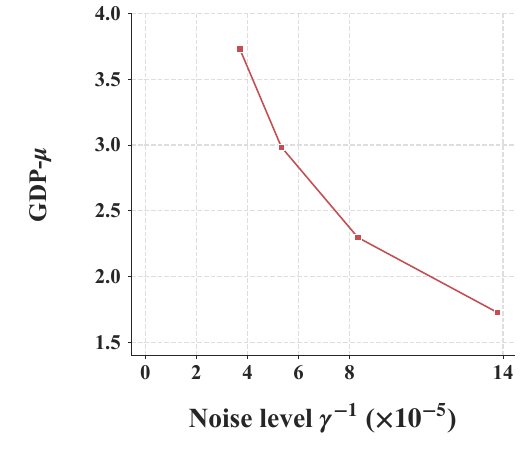}
\caption{Prediction results for the ACTG-175 data across different
noise levels \(\gamma^{-1}\). The left, middle, and right panels report
test negative log-likelihood (NLL), test root mean squared error
(RMSE), and the asymptotic GDP parameter $\mu$ corresponding to
Theorem~\ref{thm:fdp_privacy}(ii), respectively. The noise-free
setting \(\gamma^{-1}=0\) is the matched clipped-SGD baseline and is
omitted from the privacy panel.}
\label{fig:actg175}
\end{figure}

Figure~\ref{fig:actg175} shows similar mean predictive metrics across the reported moderate-noise settings.  The baseline with $\gamma^{-1}=0$ has mean NLL
\(1.1329\) and mean RMSE \(0.7491\). At
$\gamma^{-1}\approx8.32\times10^{-5}$ and $1.38\times10^{-4}$, the
mean RMSE values are \(0.7464\) and \(0.7466\), respectively. These mean RMSE values are slightly below the baseline, alongside a lower reported asymptotic GDP parameter $\mu$.

\subsection{The MNIST Dataset}
\label{subsec:mnist}

We next consider the MNIST handwritten-digit dataset of
\citet{lecun1998gradient}.\footnote{\url{https://yann.lecun.org/exdb/mnist/}}
The dataset contains 70,000 grayscale images of handwritten digits
from 0 to 9. Each image consists of \(28\times28\) pixels. The
standard split contains 60,000 training images and 10,000 test images.
MNIST is widely used to evaluate privacy-preserving neural-network
training \citep{abadi2016deep}. In this experiment, we draw a
class-stratified sample of 3,000 images from the standard training set
and retain all 10,000 test images for evaluation. The sampling scheme
preserves the original class proportions and gives approximately 300
training images for each digit. Pixel intensities are rescaled to
\([0,1]\). The reported privacy calculation concerns the selected 3,000-image training dataset.

We use the convolutional neural-network architecture considered by
\citet{li2016preconditioned}. The network contains two
convolution--max-pooling blocks with 32 and 64 filters of size
\(5\times5\). These are followed by two fully connected ReLU layers
of width 200 and a 10-class output layer. The resulting network
contains 299,306 trainable parameters. Training uses mini-batches of
size 64 and a per-sample gradient-clipping threshold of $\tau = 5 < 2\sqrt{\log(3000)}$.
We again use the terminal-decay schedule,
\(\eta_k=10^{-3}(1-k/K)^{0.501}\). For each test image, we average
the predicted class probabilities over the network states collected
during the last 20\% of the training trajectory. Performance is
evaluated using test negative log-likelihood and test classification
accuracy. The experiment is repeated independently 20 times.

\begin{figure}[!htbp]
\centering
\includegraphics[width=0.6\textwidth]
{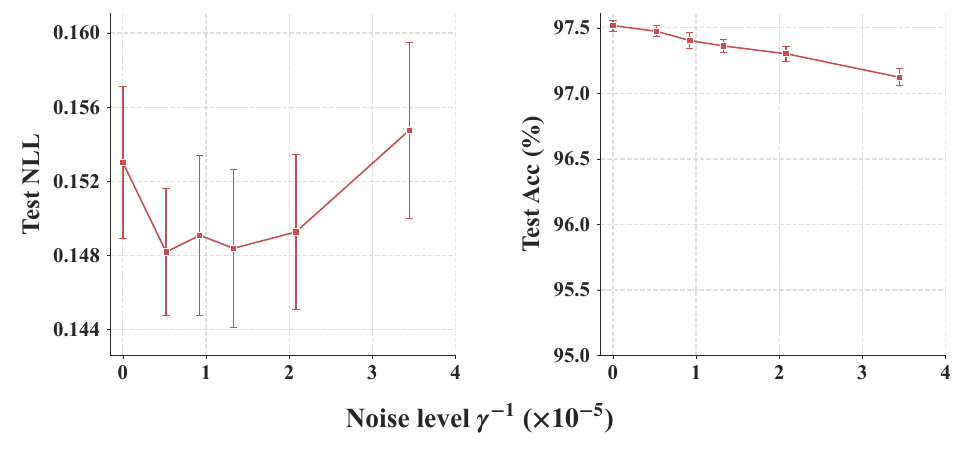}
\includegraphics[width=0.33\textwidth]
{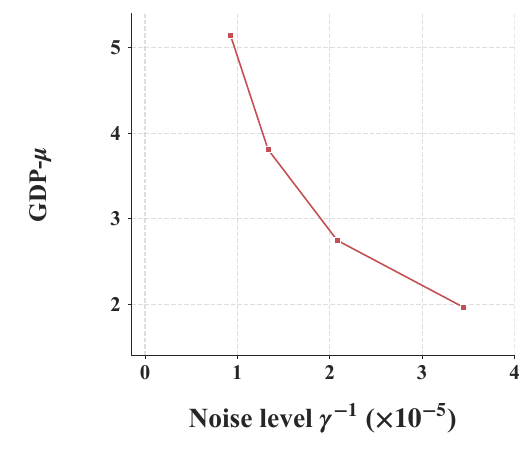}
\caption{Prediction results for MNIST across different noise levels
\(\gamma^{-1}\). The left, middle, and right panels report test
negative log-likelihood (NLL), test classification accuracy, and the
asymptotic GDP parameter $\mu$ corresponding to
Theorem~\ref{thm:fdp_privacy}(ii), respectively. The noise-free
setting \(\gamma^{-1}=0\) is the matched clipped-SGD baseline and is
omitted from the privacy panel.}
\label{fig:mnist-privacy}
\label{fig:mnist-gdp}
\end{figure}

Figure~\ref{fig:mnist-privacy} shows mean test accuracies above $97\%$ across the displayed noise levels. The baseline with
$\gamma^{-1}=0$ has mean NLL \(0.1530\) and mean accuracy \(97.518\%\).
At $\gamma^{-1}\approx2.08\times10^{-5}$ and $3.45\times10^{-5}$,
the mean accuracies are \(97.303\%\) and \(97.123\%\), respectively.
At these two noise levels, mean accuracy is approximately $0.22$ and $0.40$ percentage points below the baseline, respectively, alongside lower reported asymptotic GDP parameters.
Across the displayed moderate-noise settings, the two applications show modest changes in mean predictive metrics alongside decreasing asymptotic GDP parameters.

\section{Conclusion}
\label{sec:con}
This paper develops a finite-time analysis of SGLD for deep
nonparametric \(M\)-estimation, with guarantees for both statistical
risk and differential privacy. By treating the final-iterate law as an
algorithm-dependent PAC--Bayes posterior, the analysis separates
network approximation, the pathwise KL cost, and the finite-time
empirical optimization error. Under the stated conditions, the SGLD
output based on a suitably chosen ReLU architecture attains the
optimal rate for H\"older-smooth target functions in \(M\)-regression up to logarithmic factors. For
clipped SGLD, we establish an exact finite-sample \(f\)-DP guarantee
 and an asymptotic GDP characterization
under heterogeneous step-size schedules.  Extensive numerical studies demonstrate the satisfactory empirical
performance of SGLD across different noise levels, while the real-data
analyses illustrate the trade-off between privacy and utility for clipped
SGLD at the reported asymptotic GDP levels.

An important direction for future research is to extend SGLD to
settings with group structure or distributional heterogeneity.
Different populations may share common components while differing in
smoothness, noise level, sample size, or privacy requirements.
Recent
studies on federated SGLD and federated averaging Langevin dynamics show
that such heterogeneity can affect gradient accuracy and convergence
\citep{elmekkaoui2021federated,deng2024convergence}. Building on these
results, one could develop hierarchical or multitask SGLD procedures
with group-adaptive clipping and noise calibration, together with risk
bounds that distinguish shared and group-specific errors.

\bibliographystyle{chicago}
\bibliography{ref}

\end{document}